\documentclass[aps,preprint,nofootinbib]{revtex4}%
\usepackage{amsfonts}
\usepackage{amsmath}
\usepackage{amssymb}
\usepackage{graphicx}%
\providecommand{\U}[1]{\protect\rule{.1in}{.1in}}
\begin{document}
\preprint{ }
\title[Short title for running header]{Comment on \textquotedblleft Arnowitt--Deser--Misner representation and
Hamiltonian analysis of covariant renormalizable gravity\textquotedblright\ by
M. Chaichian, M. Oksanen, A. Tureanu}
\author{N. Kiriushcheva}
\email{nkiriush@uwo.ca}
\author{P. G. Komorowski}
\email{pkomoro@uwo.ca}
\author{S. V. Kuzmin}
\email{skuzmin@uwo.ca}
\affiliation{The Department of Applied Mathematics, The University of Western Ontario,
London, Ontario, N6A 5B7, Canada}

\begin{abstract}
The partial Hamiltonian analysis of the actions presented in the paper by M.
Chaichian, M. Oksanen, A. Tureanu (\textit{Eur. Phys. J. C 71, 1657 (2011)})
is incorrect; the true algebra of constraints differs from what they claim for
their choice of momentum constraint. Our blind acceptance of the correctness
of their constraint algebra led us to conclude, wrongly, that a few of the
models presented by the authors (sharing the same constraint algebra) are not
invariant under spatial diffeomorphism. We \textquotedblleft
proved\textquotedblright\ this by using Noether's second theorem (see first
version of the paper), but we then found a mistake in our calculations. The
differential identity of spatial diffeomorphism is intact, therefore, their
actions are invariant; but in this case, the spatial diffeomorphism gauge
symmetry cannot be compatible with their algebra. We now explicitly
demonstrate that the actual algebra of constraints is different, and briefly
describe how it affects the generator and gauge transformations of the fields. 

\end{abstract}
\volumeyear{year}
\volumenumber{number}
\issuenumber{number}
\eid{identifier}
\maketitle

The Hamiltonian analysis (partial) of the modified Ho\v{r}ava-type models was
recently considered in \cite{COTEPJC} and \cite{MasudCGQ,MasudPRD}. After
elimination of the second-class constraints and the corresponding variables
(three variables in \cite{COTEPJC} or one variable in \cite{MasudCGQ,MasudPRD}%
) the total Hamiltonian\footnote{Instead of using undetermined Lagrange
multipliers in front of the primary constraints in (\ref{eqnC61}), we write
the undetermined velocities as they appear in the Legendre transformation. In
the case under consideration this difference is not important, but for some
formulations it is crucial for the restoration of the gauge transformations of
all fields.} was obtained:%
\begin{equation}
H_{T}=\dot{N}\pi+\int dx\dot{N}_{i}\pi^{i}+N\int dxH_{0}+\int dxN_{i}%
H^{i}\text{.}\label{eqnC61}%
\end{equation}

Note that the independent variables (and conjugate momenta) of these
formulations are $N,N_{i},g_{pq}$ $(\pi,\pi^{i},\pi^{pq})$ and scalar fields,
e.g. $A\left(  \pi_{A}\right)  $ \textit{et cetera}. The closure of the
constraint algebra in the projectable case, $N=N\left(  t\right)  $, is
claimed (see Eqs. (3.23)-(3.25) of \cite{COTEPJC}, Eq. (43) of \cite{MasudCGQ}%
, and Eq. (101) of \cite{MasudPRD}) to be in the form:%

\begin{equation}
\left\{  \Phi_{0},\Phi_{0}\right\}  =0, \label{eqnC6}%
\end{equation}

\begin{equation}
\left\{  \Phi_{S}\left(  \xi_{i}\right)  ,\Phi_{0}\right\}  =0, \label{eqnC7}%
\end{equation}

\begin{equation}
\left\{  \Phi_{S}\left(  \xi_{i}\right)  ,\Phi_{S}\left(  \eta_{j}\right)
\right\}  =\Phi_{S}\left(  \xi^{j}\partial_{j}\eta_{i}-\eta^{j}\partial_{j}%
\xi_{i}\right)  \approx0,\label{eqnC8}%
\end{equation}
where%

\begin{equation}
\Phi_{0}=\int dxH_{0},\text{ \ \ \ \ \ \ }\Phi_{S}\left(  \xi_{i}\right)
=\int dx\xi_{i}H^{i}. \label{eqnC9}%
\end{equation}

Equations (\ref{eqnC6})-(\ref{eqnC8}) are similar to those presented by
Ho\v{r}ava \cite{HoravaJHEP2009}, which are\ based on the known results for
General Relativity (GR) in ADM variables with the additional projectability
condition, $N=N\left(  t\right)  $; but in the Ho\v{r}ava paper
\cite{HoravaJHEP2009}, the constraint algebra (taken from the Hamiltonian
formulation of GR in ADM variables) was written for a different choice of
shift variable $N^{k}$, and its conjugate momentum $\pi_{k}$ (primary
constraint), and its time development leads to the momentum constraint $H_{k}$
, which is related to $H^{i}$ of \cite{COTEPJC,MasudCGQ,MasudPRD} by%

\begin{equation}
H_{k}=g_{ik}H^{i}. \label{eqnC60}%
\end{equation}

This relationship (see Eq. (\ref{eqnC60})) is different from that for ordinary
field theory where the indices are raised and lowered by the Minkowski tensor
with no effect on the Poisson Brackets (PBs). Further, $\xi_{i}$ in
\cite{COTEPJC,MasudCGQ,MasudPRD} is a test function, which is just a different
form of presentation of the PBs of two constraints (e.g. see section 3 of
\cite{EPJC}), that avoids the appearance of delta functions and their
derivatives. The test functions are assumed to have a zero PB with all
variables presented in the constraints, but this property would be destroyed
if one were to use $g_{ik}\xi^{k}$ instead of $\xi_{i}$.

The Ho\v{r}ava algebra is%

\begin{equation}
\left\{  \Phi_{0},\Phi_{0}\right\}  =0,\text{ \ \ }\left\{  \Phi_{S}\left(
\xi^{i}\right)  ,\Phi_{0}\right\}  =0,\text{\ \ } \label{eqnC25}%
\end{equation}

\begin{equation}
\left\{  \Phi_{S}\left(  \xi^{i}\right)  ,\Phi_{S}\left(  \eta^{j}\right)
\right\}  =\Phi_{S}\left(  \xi^{j}\partial_{j}\eta^{i}-\eta^{j}\partial_{j}%
\xi^{i}\right)  \approx0,\text{ \ }\label{eqnC65}%
\end{equation}
where%

\[
\Phi_{0}=\int dxH_{0},\text{ \ \ \ \ \ \  \ }\Phi_{S}\left(  \xi^{k}\right)
=\int dx\xi^{k}H_{k}.
\]

It is well known that if some constraints are linear combinations of others,
with field-dependent coefficients (fields that are canonically conjugate to
variables presented in constraints, i.e. as (\ref{eqnC60})), then the
constraint algebra cannot be the same. Such a change cannot affect closure of
the algebra, but the form of closure must be different. Therefore, from a
comparison of the two algebras, one ought to conclude that spatial
diffeomorphism is not the gauge symmetry of the models in
\cite{COTEPJC,MasudCGQ,MasudPRD}.

The first version of this comment was based on this observation, and we had
decided to show, using Noether's second theorem, that there is no spatial
diffeomorphism gauge symmetry for actions that lead to (\ref{eqnC6}%
)-(\ref{eqnC8}). But after the previous version appeared on arXiv, we found
that the contribution that we claimed destroys the Differential Identity (DI)
of spatial diffeomorphism can\ in fact be compensated. All our attempts to
rescue our conclusion and to find another term did not succeed. Therefore, we
returned to the Hamiltonian formulation of \cite{COTEPJC}, since the only way
to reconcile our corrected Lagrangian analysis with the Hamiltonian one was to
investigate the possibility that the algebra of constraints (\ref{eqnC6}%
)-(\ref{eqnC8}) stated in \cite{COTEPJC,MasudCGQ,MasudPRD} is incorrect. Both
approaches, Lagrangian and Hamiltonian, must give the same result; and if the
DI of spatial diffeomorphism is valid, then spatial diffeomorphism is the
gauge symmetry, but in this case algebra (\ref{eqnC6})-(\ref{eqnC8}) must be different.

To check this algebra, all variables that correspond to secondary constraints
must be eliminated; this process makes the Hamiltonian analysis of
\cite{COTEPJC} complicated, and unfortunately the Hamiltonian constraint was
not written explicitly in the paper. Therefore we switch to another simpler
model of the authors, discussed in \cite{MasudCGQ}, in which only one field
was eliminated by solving a simple equation. For action Eq. (27)
\cite{MasudCGQ}, the total Hamiltonian (\ref{eqnC61}) has the following
explicit form of the Hamiltonian and momentum constraints (see Eq. (34) of
\cite{MasudCGQ} or Eq. (96) of \cite{MasudPRD}):%

\begin{equation}
H_{0}=\frac{1}{\sqrt{g}}\left[  \frac{1}{B}\left(  g_{ik}g_{jl}\pi^{ij}%
\pi^{kl}-\frac{1}{3}g_{ij}g_{kl}\pi^{ij}\pi^{kl}\right)  -\frac{1}{3\mu}%
g_{pq}\pi^{pq}\pi_{B}-\frac{1-3\lambda}{12\mu^{2}}B\pi_{B}^{2}\right]
\label{eqnC62}%
\end{equation}

\[
+\sqrt{g}\left[  B\left(  E^{ij}G_{ijkl}E^{kl}+A\right)  -F\left(  A\right)
+2\mu g^{ij}\nabla_{i}\nabla_{j}B\right]  ,
\]

\begin{equation}
H^{i}=-2\partial_{j}\pi^{ij}-g^{ij}\left(  2\partial_{k}g_{jl}-\partial
_{j}g_{kl}\right)  \pi^{kl}+g^{ij}\partial_{j}B\pi_{B}\label{eqnC63}%
\end{equation}
where, unlike the complicated Eqs. (3.19)-(3.20) of \cite{COTEPJC}, one has to
substitute $A=\tilde{A}\left(  B\right)  $, which results from the elimination
of the second-class constraints (see section 2.4 of \cite{MasudPLB}%
)\footnote{To simplify the notation in this comment we omit the superscript
$^{\left(  3\right)  }$, i.e. $g_{km}^{\left(  3\right)  }=g_{km}$,
$\sqrt{g^{\left(  3\right)  }}=\sqrt{g}$ .}.

Because the expression for the Hamiltonian constraint is complicated, even for
this model, we begin with a calculation of (\ref{eqnC7}) for a particular term
(second in (\ref{eqnC62})),%

\begin{equation}
\left\{  \left(  -2\partial_{j}\pi^{ij}-g^{ij}\left(  2\partial_{k}%
g_{jl}-\partial_{j}g_{kl}\right)  \pi^{kl}+g^{ij}\partial_{j}B\pi_{B}\right)
\left(  \mathbf{x}\right)  ,\int d^{3}\mathbf{y}\left(  -\frac{1}{3\mu
}\right)  \left(  \frac{1}{\sqrt{g}}g_{pq}\pi^{pq}\pi_{B}\right)  \left(
\mathbf{y}\right)  \right\}  ,\label{eqnC50}%
\end{equation}
which obviously produces non-zero contributions. The simplest non-zero
contribution of (\ref{eqnC50}) is%

\begin{equation}
-\frac{1}{3\mu}\left(  \partial_{j}B\pi_{B}\right)  \left(  \mathbf{x}\right)
\int d^{3}\mathbf{y}\left[  \left\{  g^{ij}\left(  \mathbf{x}\right)
,\pi^{pq}\left(  \mathbf{y}\right)  \right\}  \left(  \pi_{B}\frac{1}{\sqrt
{g}}g_{pq}\right)  \left(  \mathbf{y}\right)  \right]  =\frac{1}{3\mu}\frac
{1}{\sqrt{g}}g^{ij}\partial_{j}B\pi_{B}^{2}\left(  \mathbf{x}\right)  ,
\label{eqnC51}%
\end{equation}
which is proportional to $\pi_{B}^{2}$.

All other contributions that originate from (\ref{eqnC50}) are linear in
momentum, $\pi_{B}$, and cannot compensate the term in the right-hand side of
(\ref{eqnC51}). There is only one term in the Hamiltonian constraint (third
term of (\ref{eqnC62})) which is a possible source of contribution
(\ref{eqnC51}), but it enters with a different numerical coefficient.
Contribution (\ref{eqnC51}) can be compensated, in principle, but it would
unavoidably lead to a restriction on the coefficient (i.e. a particular
relationship between $\mu$ and $\lambda$). But the PB of the momentum
constraint with the third term of (\ref{eqnC62}) is%

\begin{equation}
\left\{  H^{i}\left(  \mathbf{x}\right)  ,\int d^{3}\mathbf{y}\left(
-\frac{1-3\lambda}{12\mu^{2}}\right)  \left(  \frac{1}{\sqrt{g}}B\pi_{B}%
^{2}\right)  \left(  \mathbf{y}\right)  \right\}  =0,\label{eqnC53}%
\end{equation}
and compensation of (\ref{eqnC51}) is impossible, even with a restriction on
the parameters; therefore, the claim made by the authors that PB (\ref{eqnC7})
is zero for their choice of variables ($N_{i}$ and the corresponding
momentum\ constraint $H^{i}$ (\ref{eqnC63})) is incorrect.

The structure of contribution (\ref{eqnC51}) precludes finding a result that
is proportional to the Hamiltonian constraint, and to have closure on the
secondary constraint, only a proportionality to the momentum constraint can be
expected. Indeed, the calculation of all contributions of (\ref{eqnC50})
provides such a result%

\begin{equation}
\left\{  H^{i}\left(  \mathbf{x}\right)  ,\int d^{3}\mathbf{y}\left(
-\frac{1}{3\mu}\right)  \left(  \frac{1}{\sqrt{g}}g_{pq}\pi^{pq}\pi
_{B}\right)  \right\}  =\frac{1}{3\mu}\frac{\pi_{B}}{\sqrt{g}}H^{i}.
\label{eqnC57}%
\end{equation}

Of course all terms of (\ref{eqnC62}) should be used, but even if there is
only one non-zero contribution (\ref{eqnC57}) and there is closure, the
generator built as in \cite{Castellani} for GR in ADM variables (all
first-class constraints and their algebra are needed) will be modified, and
the transformations will be different. In fact, we found additional non-zero
contributions for PB (\ref{eqnC7}), and all of them are proportional to the
momentum constraint (\ref{eqnC63}). The PB among two momentum constraints
(\ref{eqnC8}) is also different from that claimed in the papers
\cite{COTEPJC,MasudCGQ,MasudPRD,MasudPLB}.

But because the constraint algebra given in
\cite{COTEPJC,MasudCGQ,MasudPRD,MasudPLB} is incorrect, it is natural to
expect that when the standard choice of shift function, $N^{k}$, with the
primary constraint $\pi_{k}$, and the momentum constraint%

\begin{equation}
H_{k}=-g_{ki}2\partial_{j}\pi^{ij}-\left(  2\partial_{m}g_{kl}-\partial
_{k}g_{ml}\right)  \pi^{ml}+\partial_{k}B\pi_{B},\label{eqnC26}%
\end{equation}
are used, the Ho\v{r}ava algebra (\ref{eqnC25})-(\ref{eqnC65}) would follow.
We also note that calculations with (\ref{eqnC26}) are much simpler to perform
than with (\ref{eqnC63}). The PBs for the parts we considered (i.e.
(\ref{eqnC53}) and (\ref{eqnC57})) for the constraint (\ref{eqnC26}) are just
zero and the rest of the terms in (\ref{eqnC62}) also give zero. The PB of two
momentum constraints (\ref{eqnC26}) is also very easy to calculate by using
the known result for GR in ADM variables, due to the decoupling of the
additional fields%

\[
H_{k}=H_{k}^{GR}+\partial_{k}B\pi_{B},\text{ \ \ }\left\{  H_{k}^{GR}%
,\partial_{k}B\pi_{B}\right\}  =0.
\]
So for the standard choice ($N^{k}$,$\pi_{k}$,$H_{k}$), the Ho\v{r}ava algebra
(\ref{eqnC25})-(\ref{eqnC65}) follows.

To restore the transformations of all fields, as in GR, the generator is built
from all of the first-class constraints, and it is at this point that the
entire constraint algebra enters the game. For GR in ADM variables the
generator is known (see \cite{Castellani} Eq. (29)), and it needs slight
modification (taking into account the projectability condition)%

\begin{equation}
G\left(  \xi^{k}\right)  =\int d^{3}x\left[  \dot{\xi}^{k}\pi_{k}+\xi
^{k}\left(  H_{k}+\partial_{k}N^{j}\pi_{j}+\partial_{j}\left(  N^{j}\pi
_{k}\right)  \right)  \right]  . \label{eqnC80}%
\end{equation}

This is the generator of the gauge transformations that corresponds to the
primary first-class constraint $\pi^{k}$ and gives transformations of
\textit{all} fields\footnote{Here, a different convention is followed. To us,
this choice seems more natural, and it corresponds to \cite{MasudCGQ}; but in
Castellani's paper \cite{Castellani} a different order is used, and this is
why he has a minus sign on the right-hand side of his Eq. (29), compared with
our (\ref{eqnC80}).}%
\begin{equation}
\delta field=\left\{  G,field\right\}  \label{eqnC67}%
\end{equation}
With generator (\ref{eqnC80}), we find the following gauge transformations:%

\[
\delta N=0,
\]

\begin{equation}
\delta N^{i}=-\dot{\xi}^{i}-\xi^{k}\partial_{k}N^{i}+\partial_{j}\xi^{k}N^{j},
\label{eqnC30}%
\end{equation}

\[
\delta g_{pq}=-g_{qm}\partial_{p}\xi^{m}-g_{pm}\partial_{q}\xi^{m}-\xi
^{j}\partial_{j}g_{pq},
\]

\begin{equation}
\delta B=-\xi^{i}\partial_{i}B, \label{eqnC44}%
\end{equation}

\begin{equation}
\delta A=-\xi^{i}\partial_{i}A. \label{eqnC45}%
\end{equation}

For the field $A$ (not presented in generator (\ref{eqnC80})) the
transformation is obtained as follows: one must return to the eliminated
second-class constraints (see section 2.4 of \cite{MasudPLB}), which were
solved for $A$ as some function of $B$ alone; because of this condition, the
explicit form is irrelevant. Let us consider $A=\tilde{A}\left(  B\right)  $
-- the transformation is $\delta A=\delta\tilde{A}\left(  B\right)
=\frac{\delta\tilde{A}\left(  B\right)  }{\delta B}\delta B=-\frac
{\delta\tilde{A}\left(  B\right)  }{\delta B}\xi^{i}\partial_{i}B$, and the
derivatives of the same equation are $\partial_{i}A=\partial_{i}\tilde
{A}\left(  B\right)  =\frac{\delta\tilde{A}\left(  B\right)  }{\delta
B}\partial_{i}B$; the combination of these two results leads to (\ref{eqnC45}%
). Transformation (\ref{eqnC30}) is the same as that given in Eq. (41) of
\cite{KKK-3} for $\partial_{k}N=0$ (projectable case) and $\xi^{0}=0$.

Note: to derive the generator (\ref{eqnC80}) (see \cite{Castellani}) the PB of
$H_{k}$ with the total Hamiltonian is needed, or $\left\{  H_{i},\int
dyN^{i}H_{i}\right\}  $ and $\left\{  H_{i},\int dyH_{0}\right\}  $. These
results are encoded in the constraint algebra, in particular, using
(\ref{eqnC65})%

\begin{equation}
\left\{  \Phi_{S}\left(  \xi^{i}\right)  ,\Phi_{S}\left(  \eta^{j}\right)
\right\}  =\left\{  \int dx\xi^{k}H_{k},\int dy\eta^{i}H_{i}\right\}  =\int
dx\left(  \xi^{j}\partial_{j}\eta^{i}-\eta^{j}\partial_{j}\xi^{i}\right)
H_{i}~,\label{eqnC70}%
\end{equation}

\begin{equation}
\int dx\xi^{k}\left\{  H_{k},\int dy\eta^{i}H_{i}\right\}  =\int dx\xi
^{k}\left(  \partial_{k}\eta^{j}H_{j}+\partial_{j}\left(  \eta^{j}%
H_{k}\right)  \right)  ;\label{eqnC71}%
\end{equation}
and by putting $\eta^{j}=N^{j}$, we then obtain%

\[
\left\{  H_{k},\int dyN^{i}H_{i}\right\}  =\partial_{k}N^{j}H_{j}+\partial
_{j}\left(  N^{j}H_{k}\right)  .
\]

Let us build the generator for the choice of variables in
\cite{COTEPJC,MasudCGQ,MasudPRD,MasudPLB} under the assumption that the
constraint algebra (\ref{eqnC6})-(\ref{eqnC8}) presented in papers
\cite{COTEPJC,MasudCGQ,MasudPRD,MasudPLB} is correct. Calculations for
(\ref{eqnC8}) (similar to (\ref{eqnC70})-(\ref{eqnC71})) for the choice of
$N_{i}$, lead to the PB%

\[
\left\{  H^{k},\int dyN_{i}H^{i}\right\}  =g^{jk}\partial_{j}N_{i}%
H^{i}+\partial_{j}\left(  g^{ji}N_{i}H^{k}\right)
\]
and to the generator%
\begin{equation}
G\left(  \xi_{k}\right)  =\int d^{3}x\left[  \dot{\xi}_{k}\pi^{k}+\xi
_{k}\left(  H^{k}+g^{mk}\partial_{m}N_{j}\pi^{j}+\partial_{j}\left(  N^{j}%
\pi^{k}\right)  \right)  \right]  .\label{eqnC37}%
\end{equation}

Using (\ref{eqnC67}) with generator (\ref{eqnC37}), one obtains the gauge transformations:%

\begin{equation}
\delta N=0, \label{eqnC40}%
\end{equation}

\begin{equation}
\delta N_{i}=-\dot{\xi}^{m}g_{im}-\xi^{m}\dot{g}_{im}-\xi^{m}\partial_{m}%
N_{i}+\partial_{j}\left(  g^{jm}N_{m}g_{ni}\xi^{n}\right)  , \label{eqnC41}%
\end{equation}

\begin{equation}
\delta g_{pq}=-g_{qm}\partial_{p}\xi^{m}-g_{pm}\partial_{q}\xi^{m}-\xi
^{j}\partial_{j}g_{pq}, \label{eqnC43}%
\end{equation}

\[
\delta B=-\xi^{i}\partial_{i}B.
\]

Note that after the calculation ($\xi_{k}$ in (\ref{eqnC37})) one has to
substitute $\xi_{k}=g_{ki}\xi^{i}$ (for the standard choice of variables and
generator (\ref{eqnC80}), $\xi^{i}$ automatically appears). Transformation
(\ref{eqnC41}) is not a spatial diffeomorphism transformation, and this fact
might explain why the Hamiltonian formulation in
\cite{COTEPJC,MasudCGQ,MasudPRD,MasudPLB} is incomplete; and why the
transformations for lapse and shift were not reported. The situation is worse
if the true closure for (\ref{eqnC7})-(\ref{eqnC8}) is used to construct the
corresponding generator. Generator (\ref{eqnC37}) becomes much more
complicated, e.g. term (\ref{eqnC57}) leads to a contribution to the generator,%

\begin{equation}
G^{^{\prime}}\left(  \xi_{k}\right)  =\int d^{3}x\left[  \dot{\xi}_{k}\pi
^{k}+\xi_{k}\left(  H^{k}+\frac{1}{3\mu}\frac{\pi_{B}}{\sqrt{g}}\pi
^{i}+F\left(  \pi^{i}\right)  \right)  \right]  . \label{eqnC85}%
\end{equation}

Just this one extra term, which is explicitly written in Eq. (\ref{eqnC85}),
modifies the transformations of fields $N_{i}$, $B$, and $\pi^{pq}$; and the
result not only differs from spatial diffeomorphism, but the primary
constraint, $\pi^{i}$, will appear in the transformations of $B$ and $\pi
^{pq}$ (see (\ref{eqnC67})), making it impossible to find the gauge
transformations in configurational space.

This failure is related to the field-parametrisation dependence of the Dirac
method \cite{KKK-5}. For a different parametrisation (choice of independent
variables and especially \textquotedblleft primary\textquotedblright%
\ variables, i.e. those for which the conjugate momenta are primary
first-class constraints) a different gauge invariance follows. For example,
the Hamiltonian formulation of the Einstein-Hilbert action in the original
variables, metric, leads to full diffeomorphism in the formulation of either
Pirani-Schild-Skinner (PSS) \cite{PSS,PLA} or Dirac \cite{Dirac,Myths}, but in
ADM variables the transformations are different \cite{Myths} (for connection
with Lagrangian methods see \cite{KKK-1,KKK-3}). The difference in these gauge
transformations is caused by the non-canonical relationship of the two sets of
variables; but for PSS and Dirac, the result is the same since they are
canonically related and have the same algebra of constraints \cite{FKK}. Some
mechanical (finite dimensional) models are known, in which one cannot restore
the gauge invariance in the Hamiltonian formulation. These cases are often
presented as so-called counterexamples for the Dirac conjecture
\cite{Diracbook}, but all such counterexamples can be explained by a change of
field parametrisation. The choice of variables in
\cite{COTEPJC,MasudCGQ,MasudPRD,MasudPLB} provides, to the best of our
knowledge, the first field-theoretical\textquotedblleft
counterexample\textquotedblright. Is this behaviour specific to models with
extra scalar fields and the projectability condition, or can GR in ADM
variables also exhibit problems if one were to choose the shift in the form
$N_{i}$ (we are unaware of a Hamiltonian analysis for this choice of
variables)? It is difficult to answer these questions because a new constraint
algebra must be calculated, and it is much more complicated than it would be
with the standard choice of variables. For example, even all structure
\textquotedblleft constants\textquotedblright\ of such an algebra are field
dependent. Because the change of variables $N^{i}\rightarrow N_{i}$ is not
canonical (the algebra is different), the known symmetry for the ADM
formulation cannot be expected, even if the transformations for
configurational variables could be restored.

\end{document}